\documentclass[12pt]{iopart}
\usepackage{amssymb,graphicx}

\begin{document}

\title{Quanta of Local Conformational Change: Conformons in $\alpha-$helical Proteins}

\author{Victor Atanasov}
\address{SQIG, Instituto de Telecomunica\c{c}\~oes,
Av. Rovisco Pais, P-1049-001 Lisbon,
Portugal}

\author{Yasser Omar}
\address{CEMAPRE, ISEG, Universidade T\'{e}cnica de Lisboa, P-1200-781 Lisbon and SQIG, Instituto de Telecomunica\c{c}\~oes, P-1049-001 Lisbon, Portugal}

\begin{abstract}
We propose the conformon as a quantum of local conformational change for energy transfer in $\alpha-$helical proteins. The underlying
 mechanism of interaction between the quantum of excitation and the conformational degrees of freedom is nonlinear and leads to solitary wave packets of conformational energy.  The phenomenon is specific to $\alpha-$helices and not to $\beta-$sheets in proteins due to the three strands of hydrogen bonds constituting the $\alpha-$helical backbone. 
\end{abstract}


\maketitle

\section{Introduction}

Flexibility of proteins is the feature giving rise to some of their remarkable properties   \cite{polphys}.  Even minute conformational changes can modify long-range electronic interactions in proteins and open pathways for molecular motions, provided proteins are not rigid object\cite{rigprot}.  In this context, we raise the following questions: Is the flexibility of proteins relevant for energy storage and transfer? Can local protein conformation come in discrete portions of elastic energy? Is this mechanism quantum or not?  The ubiquitous $\alpha-$helical sections of the secondary structure of proteins have already been proposed by Davydov and co-workers as a conduit for quantum coherent energy transfer  due to the interchain hydrogen bonded network which stabilizes their structure and forms a path along the $\alpha-$helix  \cite{DavidovBook}.  Here we address these questions from the stand point of measuring the conformation of the $\alpha-$helix through a set of natural geometric quantities such as curvature and torsion and their coupling to a quantum of vibrational energy present on the helix. To answer the above questions, we propose a nonlinear quantum mechanism of energy transport due to a geometry coupling effect along the hydrogen bonded lattice of $\alpha-$helical sections of proteins. But first let us introduce the basic structure of proteins.

\begin{figure}[h]
\begin{center}
\includegraphics[scale=0.55]{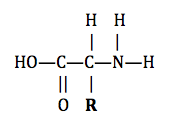}
\caption{\label{amino} Chemical structure of an amino acid, where  the symbol {\bf R} denotes the residue, a side hydrocarbon chain.}
\end{center}
\end{figure}

\begin{figure}[ht]
\begin{center}
\includegraphics[scale=0.35]{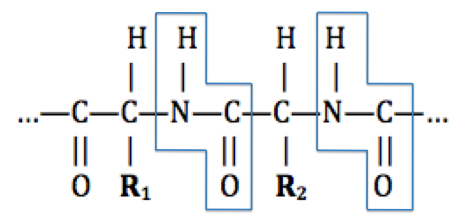}
\caption{\label{chain} A chain of amino acids, linked by peptide bonds (peptide groups are given in blue boxes), with different residues $\bf R_1$ and $\bf R_2$. Here the peptide groups are interconnected by a single C atom holding the residue. The bonds C--N and C--C can rotate, leading to a rearrangement of the chain in space.}
\end{center}
\end{figure}

\begin{figure}[ht]
\begin{center}
\includegraphics[scale=0.40]{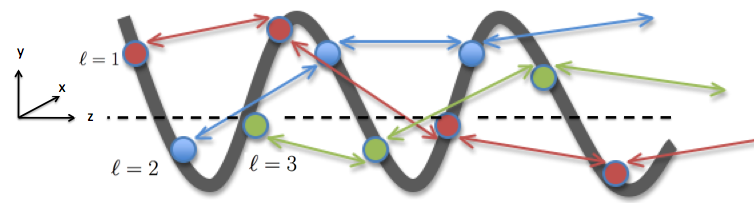}
\caption{\label{alpha} Representation of the (3D) helical structure of the  $\alpha-$helix portion of a protein. The peptide groups are represented  by circles. Hydrogen bonds  (represented by the arrows) link every third peptide group. This forms the three strands (represented in three different colours) constituting the backbone of the helix.}
\end{center}
\end{figure}

The primary structure of proteins is a chain sequence of amino acids, biomolecules which have the structure depicted in Figure \ref{amino}. Amino acids are bound together, to form  proteins, by a peptide bond constituted by the elimination of a water molecule from the acidic group and the ${\rm -NH}$ terminus. To each amino acid is attached a residue, a side hydrocarbon chain. There exist 20 different residues and their sequence gives the protein's primary structure. Apart from these residues, proteins are constituted by a repeated pattern (see Figure \ref{chain}): a group of atoms, called the peptide group (see Figure \ref{chain}). The peptide group has two functions: i.) act as a bond between carbon atoms holding the distinct residues; ii.)  due to their dipole moment make the protein fold into secondary structures stabilized by the formation of hydrogen bonds between the peptide groups. The two most common distinct shapes found in protein secondary structures are $\alpha-$helices, shown in Figure \ref{alpha}, and planar structures called $\beta-$sheets \cite{pauling}. The mechanism of energy transfer and localization in proteins proposed here concerns the $\alpha-$helical sections.

The underlying mechanism describing the local conformational dynamics of proteins for energy transfer motivated the idea  of the  {\it conformon}, a concept advanced by various authors  \cite{gree, volk, keme, scot} to play an important role in biology. It was originally envisioned as an explanation of how the energy ($\approx 0.5$ eV) released in the hydrolysis of the ATP molecule, an amount insufficient to excite the electronic structure, is  transported along biomolecules without being immediately dissipated to the  environment's degrees of freedom. This energy is slightly larger than two quanta of the Amid-I  vibration of the peptide group (Amid-I is the C=O stretch, which can drive the hydrogen bond between two peptide groups). The Amid-I vibrational quanta are thus accepted as the standard initial sinks in the biophysical processes utilizing the energy released in the hydrolysis of the ATP molecule \cite{DavidovBook}. The conformon is a concept spanning a bridge between the three physical processes in the base of the functioning of biological molecules such as proteins: i.) energy storage and transfer; ii.) conformational changes; iii.) charge transport. The conformon was introduced independently by Green and Ji  \cite{gree}, by Vol'kenstein  \cite{volk} and by Kemeny and Goklany  \cite{keme}. Green and Ji proposed the conformon as ``the free energy associated with local conformational strain."  Vol'kenstein defined it as ``the displacement of the electronic density in a macromolecule that produces the deformation of the lattice, i.e.\ the conformational strain.'' Kemeny and Goklany, using Holstein's theory  \cite{holstein},  defined the conformon as an electron spread over a molecule forming a polaron bound to a number of phonons. This bound state can then propagate from molecule to molecule.

To avoid confusion it should be mentioned that the concept of conformons as bosons of linear relaxation modes with photon statistics is also used in the context of polymer networks and dissipative liquid-like system such as glass forming fluids  \cite{Kilian}. It is interesting to note that the conformon in the definitions presented above is a quantum phenomenon and is subject to the field of quantum biophysics recently covered in  \cite{quebs}. Other conformon-like concepts in biophysics have been introduced \cite{heimburg, yakushevich} but are classical in nature.

Here we propose a physical definition of the conformon as a \textit{nonlinear} mechanism of exchange of energy and coherence between the quantum of vibrational excitation and the conformational degrees of freedom of the $\alpha-$helix which, due to the presence of the excitation, are quantized themselves. By introducing a measure for the conformation of the $\alpha-$helix, we will show that the conformon can be viewed as a quantum of local conformation.

We start the description of the conformational dynamics of biopolymers with their representation as thin rods where the relevant degrees of freedom are the torsion and the curvature. In the Kirchhoff model of an inextensible rod with fixed torsion, the curvature  dynamics allows for solitary waves in the case of non-circular cross-sections \cite{Dan*05}. A quantum particle constrained in a rod is subject to an attractive potential induced by the torsion and the curvature \cite{daco}. Applying this to the problem of
electron transport, it can be shown that this quantum potential leads to localized states, as the energy density in the
polymer is proportional to the curvature squared. This is essentially because the rod's
curvature and torsion  ``interact'' with the excitation by
inducing a potential well, which traps the electron, in addition to
creating  a nonlinearity in its Schr\"{o}dinger equation \cite{Dan*05}.
While this scenario is somewhat analogous to a  polaron (a
localized electronic bound state in a discrete lattice which is
not perfectly periodic) the origins of the two mechanisms are
quite distinct, with the curved geometry of the biopolymer playing a
key  role in  the creation of the conformon.

Our motivation for introducing the notion of the conformon is its potential for an explanation of the recently observed long-lived low frequency nonlinear vibrational states in predominantly $\alpha-$helical proteins. Pump-probe experiments revealed that low frequency nonlinear modes are essential for functionally important conformational transitions in proteins containing $\alpha-$helices \cite{Austin}. The lifetime of these states for  bacteriorhodopsin can go over 500 ps, which is intriguing considering the expected dissipation and dephasing timescales for proteins of that size. Other nonlinear mechanisms of energy transport in $\alpha-$helices, as Davydov's theory of extrinsic anharmonicity of hydrogen bonds coupled to an excitation localized on the lattice, fail to exhibit such a long lifetime \cite{scot, refhelix} considering the physical parameters of the hydrogen bonds and couplings in the models.

While the Dandoloff-Balakrishnan  model \cite{Dan*05} just discussed provides a possible dynamical
underpinning for the conformon, the aim of the present paper is to give the underlying microscopic quantum mechanical description. This task is achieved by considering the full structure of the hydrogen bonded network of the $\alpha-$helix (see Figure \ref{alpha}). In reality, $\alpha-$helices are fixed by three strands of soft hydrogen bonds stemming from the  periodically placed peptide groups. Except for reference \cite{refhelix}, previous studies simplify the description by assuming an effective single strand of hydrogen bonds, thus concentrating on the helical-symmetry preserving cases.  Here we address the symmetry breaking modes analytically, which not only have been numerically demonstrated to be less energetic and thus more stable\cite{scot, refhelix}, but also give the realistic description of the $\alpha-$helix hydrogen backbone. 
	
In the following section we construct the hamiltonian for the energy transfer along $\alpha-$helical proteins taking into account the three strands of hydrogen bonds. Next, in section \ref{sec:solution}, we solve the proposed theory. Section \ref{sec:conclusions} discusses the conformon solution and justifies it as a quantum of conformation to be found in $\alpha-$helical proteins.

\section{The Hamiltonian}

The  $\alpha-$helix of a protein contains $\nu=3.6$ peptide groups per period of the helix\cite{DavidovBook}. The equilibrium position of each peptide group is given, in a coordinate system centered with the axis of the helix (see Figure \ref{alpha}), by the radius vector
\begin{equation}\label{eq:R}
\vec{R}_{\ell} = r \left[ \vec{e}_x \cos\left(  \frac{2 \pi \ell}{\nu} \right) + \vec{e}_y \sin\left(  \frac{2 \pi \ell}{\nu} \right)   \right] + \vec{e}_z \frac{ a \, \ell}{\nu}  ,
\end{equation}
where $\vec{e}_x, \vec{e}_y$ and $\vec{e}_z$ are the orthonormal basis vectors, $\ell$ is an integer labeling each peptide group, $r=1.7\AA$ is the helix's radius and $a=5.4\AA$ is the period of the helix along $z$ \cite{pauling}. Along the $\alpha-$helix, a peptide group $\ell$ is connected to its neighbours $\ell-1$ and $\ell+1$ by rigid covalent bonds. The same peptide group $\ell$ is bound with its $\ell-3$ and $\ell+3$ neighbours through a soft hydrogen bond. These constitute the three hydrogen strands along the helix (see Figure \ref{alpha}). 

\begin{figure}[ht]
\begin{center}
\includegraphics[scale=0.45]{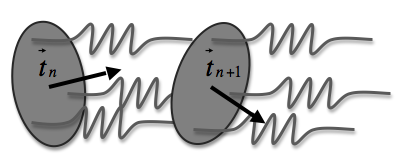}
\caption{\label{fig:gamma} Our model of the $\alpha-$helix, as a sequence of rigid disks (consisting of three consequtive peptide groups and the hydrocarbon residues attached to them) connected with three springs represented by the wavy lines corresponding to hydrogen bonds. The covalent bonds connecting and fixing the mutual orientation of two consequtive disks are treated as contraints. }
\end{center}
\end{figure}

\begin{figure}[ht]
\begin{center}
\includegraphics[scale=0.4]{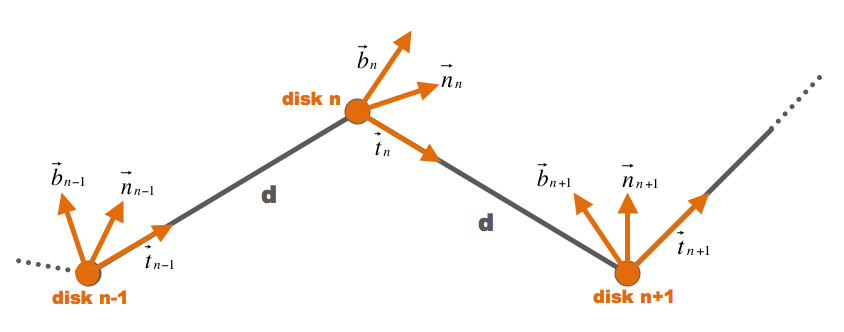}
\caption{\label{fig:curvgamma}  The discrete space curve $\Gamma$ is constructed from straight lines of constant length $d$ (called edges) along the normals to the rigid disks making up the $\alpha-$helix. A local orthonormal frame is attached to every node of $\Gamma.$ The change of the orientation of one such frame with respect to its neighbouring one along $\Gamma$ is a measure of the elastic energy stored in a particular conformation of $\Gamma$. }
\end{center}
\end{figure}

Due to the softness of the hydrogen bonds, the peptide groups can be displaced from their equilibrium positions. Let us consider sets of three consecutive peptide groups (along the helix): according to Euclidean geometry, each set will determine an unique triangle/plane containing the respective peptide groups. We can now look at the $\alpha-$ helix as a sequence of such  triangles/planes connected to each other in two different ways: i.) there is a hard covalent bond connecting the last peptide group of one triangle to the first peptide group of the next triangle; ii.) each of the three peptide groups of one triangle are connected with the peptide groups of the other triangles through the three hydrogen bond strands.  From the mechanical point of view we may treat each plane $n$, without loss of generality, as a rigid disk with an effective mass  $M_{\rm n}$ and an inertia tensor $\hat{I}_{\rm n}$ which take into account the residues attached to each peptide group.  In what follows, we will model the $\alpha-$helix as a linear chain of such rigid disks elastically connected with three springs representing the hydrogen bonds (see Figure \ref{fig:gamma}). The covalent bonds will be treated as  rigid constraints reducing extra degrees of freedom (and thereby are not depicted in Figure \ref{fig:gamma}). To quantify the local conformation in this model of the $\alpha-$helix we must introduce a measure of the  relative orientations of the disks.  To achieve this we track the change in the orientation of the unit normal vector to each disk with respect to its neighboring one. Considering the set of all those vectors, represented one after the other and separated by an effective constant distance $d$ (the $\alpha-$helix is assumed inextensible, as discussed later), we construct a discrete space curve $\Gamma$ (see Figure \ref{fig:curvgamma}). Next, we associate elastic energy to every configuration of $\Gamma,$ since every deviation from the equilibrium position either presses upon or stretches the hydrogen bonds from the three strands. This way the conformation of the $\alpha-$helix in space can be quantified.

We describe the configuration of the $\alpha-$helix as a discrete curve $\Gamma$ comprised of $N$ nodes (representing $N$ disks or $3N$ peptide groups) and $N-1$ straight edges of length $d$  together with an assignment of local material orthonormal frames in the standard  Darboux-Ribaucour notation $\vec{f}_{\rm n} = (\vec{b} _{\rm n}, \vec{n}_{\rm n}, \vec{t}_{\rm n} )^T$ per node, where $T$ denotes transposition \cite{discurv}. Here $ \vec{t}_{\rm n}$ is the unit normal vector to a  disk $\rm n$ giving the orientation of the associated edge (see Figure \ref{fig:curvgamma}).

Now, we assign an elastic energy $E(\Gamma)$ to any conformation of  $\Gamma.$ Note, we assume the $\alpha-$helix is inextensible, therefore we do not include stretching energy. It is straightforward to drop this assumption by also including a stretching term. Such a term is going to generate an acoustic mode which is not the focus of the present study. 

To compute the elastic energy $E(\Gamma)$ we use a  three dimensional symmetric\footnote{in the the material frame} strain tensor $S.$ In the first approximation, as in Hooke's law, $E(\Gamma)$  is quadratic  in the displacements from the initial equilibrium configuration $ \vec{\xi}_{\rm n}^{\;0}$ as measured by the Darboux vector $\vec{\xi}_{\rm n}.$  The Darboux vector gives the relation between $\vec{f}_{\rm n}$ and $\vec{f}_{\rm n+1}.$ The sequence of Darboux vectors $\vec{f}_{\rm 1}, \vec{f}_{\rm 2} \ldots \vec{f}_{\rm N}$ gives the evolution of $\Gamma$ \cite{Balakrishnan*93}. Finally, for the elastic energy we have:
\begin{equation}\label{Egamma}
E(\Gamma)=\frac12 \sum_{ {\rm n=1}}^N  \left( \vec{\xi}_{\rm n} - \vec{\xi}_{\rm n}^{\;0} \right)^T S \left( \vec{\xi}_{\rm n} - \vec{\xi}_{\rm n}^{\;0} \right),
\end{equation}
where $ \vec{\xi}_{\rm n}^{\;0}$ is the initial configuration of $\Gamma$ and the Darboux vector $ \vec{\xi}_{\rm n}$ in the local material frame $\vec{f}_{\rm n}$ is given by:
\begin{equation}\label{Darboux}
\vec{\xi}_{\rm n} = \kappa^n_{\rm n} \vec{n}_{\rm n} +  \kappa^g_{\rm n} \vec{b}_{\rm n}+  \tau_{\rm n} \vec{t}_{\rm n} ,
\end{equation}
where  $\kappa^g_{\rm n}$ is the geodesic curvature, $\kappa^n_{\rm n}$ is the normal curvature and $\tau_{\rm n}$ is the torsion defined on each disk of $\Gamma.$ In the continuum limit, where $N \to \infty$ and $d \to 0,$ one obtains the bending and twisting energies of the inextensible Kirchhoff rod \cite{Audoly*08}.
 
To compute the curvatures and torsion present in equation (\ref{Darboux}) we start by establishing a mapping between the local material frames $\vec{f}_{\rm n}$  and the fixed laboratory frame $\vec{e}= ( \vec{e} _{x}, \vec{e}_{y}, \vec{e}_{z} )^T.$ Then we give the orientation of each frame $\vec{f}_{\rm n}$ in terms of the Euler angles:
\begin{equation}
\vec{f}_{\rm n} = R_{\rm n} (\psi_{\rm n}, \theta_{\rm n}, \phi_{\rm n}) \, \vec{e},
\end{equation}
where $R_{\rm n}$ is the product of the three Euler matrices representing three rotations by $\phi_{\rm n}$ degrees around the $z-$axis,  by $\theta_{\rm n}$ degrees around the $x-$axis and by $\psi_{\rm n}$ degrees around the $z-$axis again. In this way, the transformation of the frame from disk to disk, i.e. the discrete analog of the Darboux-Ribaucour equations \cite{Balakrishnan*93}, are given by
\begin{equation}
\vec{f}_{\rm n+1}= R_{\rm n+1} R_{\rm n}^{-1} \vec{f}_{\rm n}, \qquad \frac{1}{d}\left( \vec{f}_{\rm n+1} - \vec{f}_{\rm n}\right)= A_{\rm n}  \vec{f}_{\rm n},
\end{equation}
where $
A_{\rm n}= (R_{\rm n+1} R_{\rm n}^{-1} - \mathbb{I})/d.
$ In the  continuous limit, when $N \to \infty$ and $d \to 0,$ the elements $(A_{\rm n})_{31}, (A_{\rm n})_{23}$ and $(A_{\rm n})_{12}$ of the matrix $A_{\rm n}$ give, respectively, the geodesic curvature,  the normal curvature and torsion of the space curve \cite{Balakrishnan*93}. We compute $A_{\rm n}$ in our discrete case   and identifying the same elements we finally obtain the coefficients of the Darboux vector in equation (\ref{Darboux}):
\begin{eqnarray}\label{kg}
\kappa^g_{\rm n} =  \sin{\psi_{\rm n}} \sin{\theta_{\rm n}} \frac{\phi_{\rm n+1} - \phi_{\rm n}}{d} + \cos{\psi_{\rm n}} \frac{\theta_{\rm n+1} - \theta_{\rm n}}{d} ,\\\label{kn}
\kappa^n_{\rm n} =    \cos{\psi_{\rm n}} \sin{\theta_{\rm n}} \frac{\phi_{\rm n+1} - \phi_{\rm n}}{d} - \sin{\psi_{\rm n}} \frac{\theta_{\rm n+1} - \theta_{\rm n}}{d}   ,\\\label{tau}
\tau_{\rm n} =   \frac{\psi_{\rm n+1} - \psi_{\rm n}}{d} + \cos{\theta_{\rm n}} \frac{\phi_{\rm n+1} - \phi_{\rm n}}{d} .
\end{eqnarray}

The interpretation of the Euler angles is the following: $\theta_{\rm n}$ and $\phi_{\rm n}$ are the polar and azimuthal angles, respectively, of $\vec{t}_{\rm n}$ in the lab frame $\vec{e}.$ The angle $\psi_{\rm n}$ rotates the $(\vec{n}_{\rm n},\vec{b}_{\rm n})$ plane around $\vec{t}_{\rm n}$ and thus leaves $\vec{t}_{\rm n}$ unchanged. This means that the normal vectors $\vec{t}_{\rm n}$ are mapped onto the $z-$axis of the lab frame through {\it precession} $\phi_{\rm n}$ around this axis and {\it tilting} $\theta_{\rm n}$ off this axis.  The differences in the polar and azimuthal angles of $\vec{t}_{\rm n}$ between two neighboring disks are not associated with the choice of a cartesian system $\vec{e}$ and are a measure of the bending and twisting of the initial conformation of $\Gamma;$
$\psi_{\rm n}$ constitutes the {\it gauge freedom} in the problem which we fix due to the existence of a covalent constraint to each disk. We set $\psi_{\rm n}$ to be a constant  for every $\rm n$ and we also fix the strain tensor 
\begin{equation}\label{s}
S=\alpha \mathbb{I}, \qquad \alpha \in \mathbb{R}
\end{equation}
in the basis $\vec{f}_{\rm n}$ corresponding to an isotropic elastic response to any change of conformation of $\Gamma.$

The total hamiltonian describing the quantum dynamics of a quantum of vibrational excitation present on the $\alpha-$helix modeled as $\Gamma$ is a collection of an elastic part $\hat{H_{\Gamma} }$ , an excitation part $\hat{H}_{ex}$  and an interaction term $ \hat{H}_{int}$ between them. A kinetic energy term of the form
$\hat{H}_{kin}=\frac12 \sum_{\rm n} \vec{\omega}_{\rm n}^{t} \hat{I}_{\rm n} \vec{\omega}_{\rm n}$ (where $\hat{I}_{\rm n}$ is the moment of inertia tensor of each disk and $\vec{\omega}_{\rm n}$ is its angular velocity) can be included but we will omit it in accordance with the adiabatic approximation. This is justified due to the enormous mass and moment of inertia tensor of the material disks. As a result the excitation's motion adjusts instantaneously to the conformation of $\Gamma,$ that is for the excitation's dynamics the slow dynamics of conformational change is effectively static.

We are now in a position to write the quantum energy operator for the elastic part of the total hamiltonian making the variables from equation (\ref{Egamma}) operator valued. Here we choose to define the initial condition for $\Gamma$ by setting $\vec{\xi}_{\rm n}^{\;0} =0$ corresponding to the equilibrium configuration of $\Gamma$ for which the protein is helical. The energy consumed in the deviation from this equilibrium state is obtained by substituting (\ref{s}) in (\ref{Egamma}) and then using (\ref{kg}), (\ref{kn}) and (\ref{tau}):  
\begin{eqnarray}\label{Helastic}
\hat{H}_{\Gamma}&=&\frac{\alpha}{2} \sum_{ {\rm n}}  \left[  (\kappa^n_{\rm n})^2 +(\kappa^g_{\rm n})^2  +  \tau_{\rm n}^2  \right]\\
\nonumber &=& \frac{\alpha}{2 d^2} \sum_{ {\rm n}}  \left[ 
(\hat{\phi}_{\rm n+1} - \hat{\phi}_{\rm n})^2 + (\hat{\theta}_{\rm n+1} - \hat{\theta}_{\rm n})^2 
\right].
\end{eqnarray}

The energy operator for the vibrational excitation present in the $\alpha-$helix $\hat{H}_{ex}$ is given in the standard manner
\begin{equation}\label{hopping}
\hat{H}_{ex} = \sum_{ {\rm n }}  \left[ E_0 \hat{B}_{\rm n}^\dag \hat{B}_{\rm n} - J ( \hat{B}_{\rm n}^\dag \hat{B}_{\rm n+1} + \hat{B}_{\rm n}^\dag \hat{B}_{\rm n-1}  ) \right],
\end{equation}
where $\hat{B}_{\rm n}^\dag$ and $\hat{B}_{\rm n}$ are the excitation's creation and annihilation operators for disk $\rm n;$ $J$ is the hopping amplitude kept constant due to the inextensibility condition on $\Gamma.$
Here, for simplicity, we assume the chain is infinite and do not consider the effect of an excitation entering the chain from either end or finite boundary conditions. 

The interaction term $\hat{H}_{int}$ between the excitation and orientational degrees of freedom is deducible from the inextensibility condition on $\Gamma$ and the local modification of the on-disk energy for the curved conformation of $\Gamma$ in the nearest neighbor approximation,  that is $E_{\rm n} \approx E_0 + {\partial E \over \partial  ({\theta}_{\rm n+1} - {\theta}_{\rm n}) } ({\theta}_{\rm n+1} - {\theta}_{\rm n}) + {\partial E \over \partial  ({\phi}_{\rm n+1} - {\phi}_{\rm n}) } ({\phi}_{\rm n+1} - {\phi}_{\rm n}).$ Finally,
\begin{eqnarray}\label{interaction}
 \hat{H}_{int} = \sum_{ {\rm n} }   \left\{  \chi_{\phi} ( \hat{\phi}_{\rm n+1} - \hat{\phi}_{\rm n})\right.\hat{B}_{\rm n}^\dag \hat{B}_{\rm n}+ \left. \chi_{\theta} (\hat{\theta}_{\rm n+1} - \hat{\theta}_{\rm n})  \hat{B}_{\rm n}^\dag \hat{B}_{\rm n}  \right\},
\end{eqnarray}
where $\chi_{\phi}$ and $\chi_{\theta}$ are coupling constants, small as compared to the conformational energy $2 d^2 \chi_{\phi}/\alpha < 1$ and $2 d^2 \chi_{\theta}/\alpha < 1.$

The total hamiltonian is the sum of (\ref{Helastic}), (\ref{hopping}) and (\ref{interaction}).  We explore the theory
\begin{equation}\label{hamiltonian}
\hat{H}=\hat{H_{\Gamma} } + \hat{H}_{ex}+ \hat{H}_{int} .
\end{equation}

\section{Solving the Theory}\label{sec:solution}

Our analysis of (\ref{hamiltonian}) will be based upon a product trial wave function
$
\left|  \Psi \rangle \right. = \left| \psi \rangle \right. \left| \phi \rangle \right.
$
in which $\left| \psi \rangle \right.$ describes a single excitation in $\Gamma$ as 
\begin{equation}\label{WF}
\left| \psi \rangle \right. = \sum_{\rm n} a_{\rm n}(t) \hat{B}_{\rm n}^\dag \left|0 \rangle_{ex} \right. ,
\end{equation}
where $\left|0 \rangle_{ex} \right.$ is the vacuum state of the its oscillators and $\left| \phi \rangle \right.$ is a coherent state for the angular displacements, for which the following hold
\begin{equation}
\left. \langle \phi \right|  \hat{\theta}_{\rm n} \left| \phi \rangle \right. = {\theta}_{\rm n}, \qquad  \left. \langle \phi \right|  \hat{\phi}_{\rm n} \left| \phi \rangle \right. = {\phi}_{\rm n}.
\end{equation}
Here $a_{\rm n}(t)$ is a complex number representing the probability amplitude for finding the excitation in a particular site;  ${\theta}_{\rm n}$ and  ${\phi}_{\rm n}$ are the average values of the angular displacements of $\Gamma$'s edges. Next we will minimize the average value of $\hat{H}$ with respect to the product wave function by calculating $ H=
\left. \langle \Psi \right|  \hat{H} \left| \Psi \rangle \right.,$
where we have to keep in mind the normalization condition
$
\sum_{\rm n} a_{\rm n} a_{\rm n}^\ast =1.
$
For the averaged energy function in the adiabatic approximation we obtain
\begin{eqnarray}
H&=& \sum_{\rm n} \left\{  E_0  a_{\rm n} a_{\rm n}^\ast - J (a_{\rm n}^\ast a_{\rm n+1} +  a_{\rm n}^\ast a_{\rm n-1} )    \right. +\\
\nonumber && +  \left[ \chi_{\phi} ({\phi}_{\rm n+1} - {\phi}_{\rm n})  + \chi_{\theta} ({\theta}_{\rm n+1} - { \theta}_{\rm n}) \right] a_{\rm n} a_{\rm n}^\ast  +\\
\nonumber && \left. + \frac{\alpha}{2 d^2} \left[  ({\phi}_{\rm n+1} - {\phi}_{\rm n})^2   +  ({\theta}_{\rm n+1} - {\theta}_{\rm n})^2 \right] \sum_{\rm m} a_{\rm m} a_{\rm m}^\ast \right\}.
\end{eqnarray}
Next we find the corresponding Hamilton's equations 
\begin{eqnarray}
i \hbar \frac{\partial  a_{\rm n}}{\partial t}&=& \left(  E_0  + W  + \Upsilon_{\rm n}  \right)a_{\rm n}  - J (a_{\rm n+1} + a_{\rm n-1} ),
\end{eqnarray}
where
$
\Upsilon_{\rm n} =\chi_{\phi} ({\phi}_{\rm n+1} - {\phi}_{\rm n})  + \chi_{\theta}  ({\theta}_{\rm n+1} - { \theta}_{\rm n})
$
and the total conformational  energy is
\begin{equation}
W= \frac{\alpha}{2 d^2} \sum_{\rm n} \left[  ({\phi}_{\rm n+1} - {\phi}_{\rm n})^2   +   ({\theta}_{\rm n+1} - {\theta}_{\rm n})^2 \right].
\end{equation}
The other  two Hamilton's equations for ${\phi}_{\rm n}$ and ${\theta}_{\rm n}$ constitute the static (since we are in the adiabatic approximation) configuration of $\Gamma:$ 
\begin{eqnarray}
{\phi}_{\rm n+1} - 2{\phi}_{\rm n} + {\phi}_{\rm n-1}= \frac{2 d^2\chi_{\phi} }{\alpha} ( \left| a_{\rm n-1} \right|^2 -   \left| a_{\rm n} \right|^2 ), \\
{\theta}_{\rm n+1} - 2{\theta}_{\rm n} + {\theta}_{\rm n-1}= \frac{2 d^2\chi_{\theta} }{\alpha} ( \left| a_{\rm n-1} \right|^2 -  \chi_2 \left| a_{\rm n} \right|^2).
\end{eqnarray}
One possible solution is 
\begin{eqnarray}\label{eq:phi_theta}
{\phi}_{\rm n} - {\phi}_{\rm n+1}=   \frac{2 d^2\chi_{\phi} \left| a_{\rm n} \right|^2}{\alpha} , \;
{\theta}_{\rm n} - {\theta}_{\rm n+1} =  \frac{2 d^2\chi_{\theta} \left| a_{\rm n} \right|^2 }{\alpha}\;
\end{eqnarray}
with 
$
\Upsilon_{\rm n} =-\upsilon \left| a_{\rm n} \right|^2$ and 
\begin{eqnarray}\label{upsilon}
\upsilon= \frac{2 d^2}{\alpha}\left( \chi_{\phi}^2    + \chi_{\theta}^2 \right).
\end{eqnarray}
A gauge transformation of the wavefunction (\ref{WF})
\begin{equation}
 a_{\rm n}(t) =  q_{\rm n}(t) \exp{ \left\{ -\frac{i}{\hbar} t  \left(E_0 + W - 2 J \right) \right\}}
\end{equation}
simplifies the equation for the excitation's amplitude to be found on the $\rm n-$th edge
\begin{eqnarray}\label{DNLS}
i \hbar \frac{\partial  q_{\rm n}}{\partial t} + J (q_{\rm n+1}  - 2 q_{\rm n} + q_{\rm n-1} )+\upsilon   \left| q_{\rm n} \right|^2 q_{\rm n}=0,
\end{eqnarray}
which is the discrete nonlinear Schr\"odinger equation (DNLS) \cite{Abl}. 
An approximate for $J/\upsilon \gg 1$ stationary (consistent with the adiabatic approximation) soliton-like solution centered at ${\rm n} = {\rm n}_0$ has the squared amplitude
\begin{equation}\label{conformonI}
\left| q_{\rm n}  \right|^2 = \frac{\upsilon}{8J} {\rm sech}^2\left[ \frac{ \upsilon (  {\rm n} - {\rm n}_0 )  }{4J}  \right].
\end{equation}
Here $q_{\rm n}$ is the probability density to find the quantum of vibrational excitation on disk $\rm n.$ Notice that the evolutionary equation which $q_{\rm n}$ obeys is the DNLS which has both continuous and discrete spectra of solutions. The solution (\ref{conformonI}) belongs to the discrete spectrum of (\ref{DNLS}) and has a bell-like shape with exponentially vanishing tails. The most probable place to find the excitation is in a region around ${\rm n}_0$ whose extent is determined by $ \upsilon/{4J},$ that is by the physical parameters in the model:  the distance $d$ between the disks, the strain tensor constant $\alpha$,  the coupling constants $\chi_{\phi}$ and $ \chi_{\theta},$  and  the hopping amplitude $J$.

Using (\ref{eq:phi_theta}) and imposing the physically consistent conditions ${d^2\chi_{\phi}}/{\alpha}\ll 1$ and $ { d^2\chi_{\theta}}/{\alpha}\ll 1$ we have the estimate
\begin{eqnarray}\label{bound_phi_theta}
0 \leq |{\phi}_{\rm n+1} - {\phi}_{\rm n} |  \ll 1, \qquad
0 \leq |{\theta}_{\rm n+1} - {\theta}_{\rm n} | \ll 1.
\end{eqnarray}
Now we turn to the corresponding conformation of $\Gamma.$ The total curvature at each disk is 
\begin{eqnarray}
\kappa_{\rm n} && = \sqrt{ (\kappa^g_{\rm n})^2 + (\kappa^n_{\rm n})^2}=\\
\nonumber &&=\frac{1}{d}\sqrt{ \sin^2{\theta_{\rm n}} (\phi_{\rm n+1} - \phi_{\rm n})^2 + (\theta_{\rm n+1} - \theta_{\rm n})^2     }
\end{eqnarray}
and the torsion having fixed the gauge freedom is 
\begin{equation}
\tau_{\rm n} = \frac{\cos{\theta_{\rm n}} ( {\phi_{\rm n+1} - \phi_{\rm n})}}{d}.
\end{equation}
Taking into account (\ref{bound_phi_theta}) we may approximate $\cos{\theta_{\rm n}}\approx 1$ and $\sin^2{\theta_{\rm n}} \approx 0$ to arrive at an approximate expression for the conformation of $\Gamma$ in terms of the excitation's presence:
\begin{eqnarray} \label{conformonII}
\nonumber \kappa_{\rm n} \approx \frac{2 d\chi_{\theta} }{\alpha} \left| q_{\rm n} \right|^2 \\
\\
\nonumber  \tau_{\rm n} \approx  -  \frac{2 d\chi_{\phi} }{\alpha} \left| q_{\rm n} \right|^2, 
\end{eqnarray}
where $\left| q_{\rm n} \right|^2$ is given by (\ref{conformonI}). The physical parameters in the model $ J, \alpha, \chi_{\theta}, \chi_{\phi} $ and $d$  establish proportionality between the material response of the $\alpha-$helix in terms of the curvature $\kappa_{\rm n}$ and the torsion $\tau_{\rm n}$  at disk $\rm n$ and the probability amplitude $ q_{\rm n}$ of the localization of the quantum of vibrational excitation on that same disk. This localization leads to an elastic response in the corresponding configuration of $\Gamma,$ with the respective energy given by (\ref{Helastic}).  Provided the five physical parameters are estimated, the conformon's energy can be determined and a proper experiment to detect it proposed. This is beyond the focus of the present study, although we have a reasonable guess for the numerical values of the parameters. Since $d$ is the distance between two adjacent disks in the model, it is of the order of the period of the $\alpha-$helix: $d \approx 0.5\;  \rm nm.$ The strain tensor constant $\alpha$ takes into account the simultaneous response of three hydrogen bonds. The spring constant $k$ for the hydrogen bond in the $\alpha-$helical sections of proteins is estimated to be $k=19.5\; \rm N/m$ \cite{scot}, leading to an estimate of $\alpha \approx 1\times 10^{-19} \; \rm eV.m^2.$\footnote{This estimate can be obtained by equating two equivalent expressions: on one side,  the elastic energy written for 3 equivalent springs for 10\% of the inter-disk distance $d$ deviation from their equilibrium position (a non-melting condition according to the empirical Lindemann criterion \cite{solidstate}) and, on the other side, the elastic energy written in terms of strain tensor formalism.}  The hopping amplitude for the Amid-I quantum of vibration along the $\alpha-$helix is of the order of $J \approx 1\times 10^{-3} \rm eV$ \cite{DavidovBook, refhelix}. The constants $\chi_{\theta}$ and $\chi_{\phi}$ coupling the conformation of $\Gamma$ to the on-disk energy are small, most likely smaller (rotational stiffness of chemical bonds is smaller that compressional stiffness) that the coupling between phonons and the Amid-I quanta ($1 \times 10^{-3}\; \rm eV$ in the case of the Davydov soliton \cite{scott}) thus rendering the estimate (\ref{bound_phi_theta}) physically correct.

\section{Conclusions}\label{sec:conclusions}

We have showed that the vibrational energy localized on the helix acts through a curvature and torsion coupling effect to distort the structure of the helix, and then the helical distortion reacts again through the curvature and torsion couplings to trap the oscillation energy and prevent dispersion. This is the self-localization or self-trapping mechanism behind the conformon's existence. Even more surprising is the observation that the curvature and torsion of the discrete space curve $\Gamma$ used to model the $\alpha-$helix  are related to the squared probability amplitude of the quantum of vibrational excitation, as expressed (\ref{conformonII}). This shows that the conformation of the $\alpha-$helix itself is quantized, justifying the conformon as a quantum of conformation.

In this paper we have investigated the role of excitation-conformation interaction on the formation of solitary wave packets of conformational energy in $\alpha-$helical proteins. The results obtained here are valid at zero temperature. However, the physical meaning of (\ref{conformonI}) and (\ref{conformonII}) is experimentally relevant. Pump-probe experiments revealed that low frequency nonlinear modes are essential for functionally important conformational transitions in proteins containing $\alpha-$helices \cite{Austin}. The characteristic lifetime of these states is 15 ps (although for the vision-relevant  bacteriorhodopsin it can go over 500 ps). The same experiments also revealed that amino acids and predominantly $\beta-$sheet proteins do not have such long-lived states. The proposed model already makes the qualitative distinction between  sections of the secondary structure of proteins due to the full use of the geometry of the hydrogen bonds' network in $\alpha-$helices. Quantitatively the model's viability would be decided upon determination of the physical constants which remains an open problem. Other nonlinear mechanisms of energy transport in $\alpha-$helical proteins fail to exhibit such a long lifetime \cite{scot, refhelix}. Thus the conformon proposed here might have the answer for the mechanism of direct coherent flow of conformational energy for a variety of vital biological processes ranging from electron transfer \cite{e-transfer} to enzyme action \cite{enzyme}. 

\section*{Acknowledgements}

The authors would like to thank the support from project IT-QuantTel, as well as from Funda\c{c}\~{a}o para a Ci\^{e}ncia e a
Tecnologia (Portugal), namely through programs POCTI/POCI/PTDC and project PTDC/EEA-TEL/103402/2008 QuantPrivTel, partially funded by FEDER (EU).

\end{document}